\documentclass{article}
\usepackage{stywhispers,amsmath,epsfig}
\usepackage{booktabs}
\usepackage[utf8]{inputenc}
\usepackage{graphicx}
\usepackage{enumitem}
\usepackage{chemformula}
\usepackage{float}

\title{Corrosion monitoring on zinc electroplated steel using shortwave infrared hyperspectral imaging}
\name{T. De Kerf $^1$,Z. Zahiri $^2$,  P. Scheunders $^2$, S. Vanlanduit $^1$ \thanks{This research is funded by the Belgian SPF Economy ETF—PhairywinD project.}}
\address{$^1$ InViLab, Faculty of applied engineering, University of Antwerp\\
	$^2$ Imec-VisionLab, physics department, University of Antwerp}
\begin{document}
%
\maketitle
\begin{abstract}
In this study, we investigate the use of hyperspectral imaging (HSI) to inspect the formation of corrosion products on galvanised carbon steel samples. Ten samples were subjected to an accelerated corrosion test with different exposure times. The analysis is performed in a two-step procedure: First, the different corrosion minerals are identified by microscopic Fourier transform infrared spectroscopy (FTIR) at specific locations on the samples. The following corrosion minerals are identified: \ch{ZnO} (zincite/zinc oxide) \ch{Zn_{5}(OH)_{8}Cl_{2}.H_{2}O} (simonkolleite), \ch{ZnCO_{3}} (smithsonite), \ch{Zn_{5}(CO _{3})_{2}(OH)_{6}} (marionite/hydrozincite). Second, the identified corrosion minerals are correlated with the HSI spectra for these specific locations. This correlation provides us with the spectra in the SWIR region and allows us to construct a classification map for the different corrosion minerals. The results show that we are able to identify the different minerals using HSI camera. This proposed methodology allows us to speed up the inspection process, compared to FTIR, while still accurately distinguishing between the different corrosion minerals. 
\end{abstract}
\begin{keywords}
Shortwave Infrared Imaging, Hyperspectral Imaging, Corrosion, Zinc Electroplated Steel, FTIR, SAM, Mineral analysis
\end{keywords}
\section{Introduction}
\label{sec:intro}
Studies have shown that corrosion costs amount to 3.4\% of global GDP \cite{Koch2017}. Common corrosion control measures, such as permanent anti-corrosion coatings or more accurate corrosion detection methods, could reduce the overall cost of corrosion. There are several methods for manipulating carbon steel samples to improve corrosion resistance. These methods include: using different alloys, applying organic coatings, powder coating the metal, and electrogalvanizing. Hot-dip galvanizing involves immersing iron or steel in a bath of molten zinc to produce a corrosion-resistant, multilayer coating of zinc-iron alloy and zinc metal. While the steel is immersed in the zinc, a metallurgical reaction occurs between the iron in the steel and the molten zinc. This zinc coating produced protects the steel in several ways:
\begin{itemize}[noitemsep,topsep=0pt]
\item The zinc coating deposited acts as a sacrificial anode. In normal environments that are corrosive to iron, the anode is attacked first, rather than iron, which acts as a cathode. 
\item The zinc coating forms a highly resistant barrier that separates the iron from harmful, corrosive environments.
\item When the zinc layer corrodes, it forms a protective layer of zinc carbonate. This additional layer covers the sample with a strongly adherent and mechanically resistant layer.
\end{itemize}

There are several methods to analyze the corrosion resistance of zinc coatings. Electrochemical Impedance Spectroscopy (EIS)\cite{Wijesinghe2017}, Scanning Vibrating Electrode Technique (SVET)\cite{Wijesinghe2017}, Fourier Transform Infrared Spectroscopy (FTIR)\cite{Kasperek1998}, Atomic Force Microscopy (AFM)\cite{Klassen2001} and Scanning Electrode Microscope (SEM)\cite{Klassen2001}. 
However, these methods are limited to single-point measurements (FTIR, SVET), are destructive methods (EIS), or are suitable only for laboratory use (AFM, SEM). Visual spectrum imaging \cite{Miyachi2021, DeKerf2021} mitigates the above drawbacks, but since the corrosion products of zinc usually have a white hue, it is difficult to achieve satisfactory detection accuracy. 

Hyperspectral imaging allows us to combine a chemical measurement approach with a non-contact measurement and image a large field of view. Identifying the corrosion minerals present in the sample can provide us with valuable information about the corrosion degradation process \cite{DeKerf2022}. This method of nondestructive evaluation could provide a rapid and noninvasive detection method for further investigation of the corrosion process in electrogalvanized coatings.

\section{Materials and Methods}

\label{sec:matmet}
\subsection{Sample preparation}
\begin{figure}[h!tpb]
    \begin{minipage}{0.32\linewidth}
        \centering
        \includegraphics[width=\linewidth,height = 8cm]{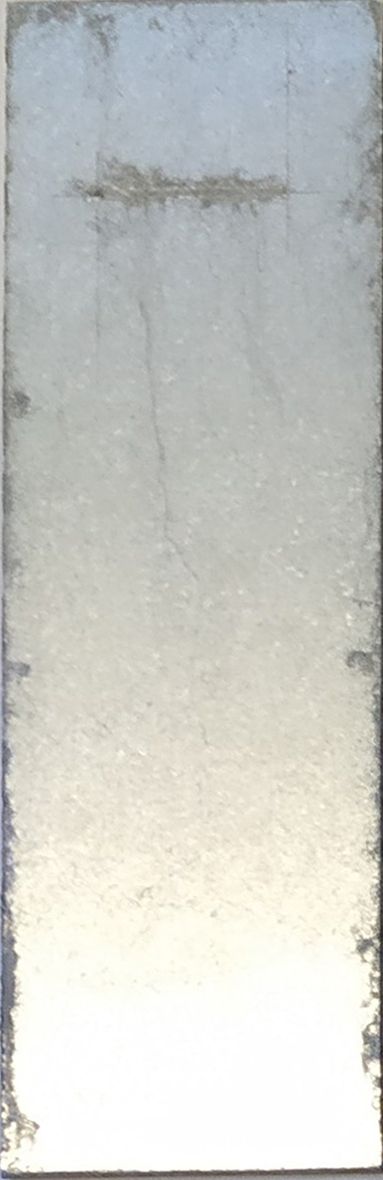} 
        \label{fig:s1_rgb}
        \centerline{(a) Sample one}

    \end{minipage}\hfill
        \begin{minipage}{0.32\linewidth}
        \centering
        \includegraphics[width=\linewidth,height = 8cm]{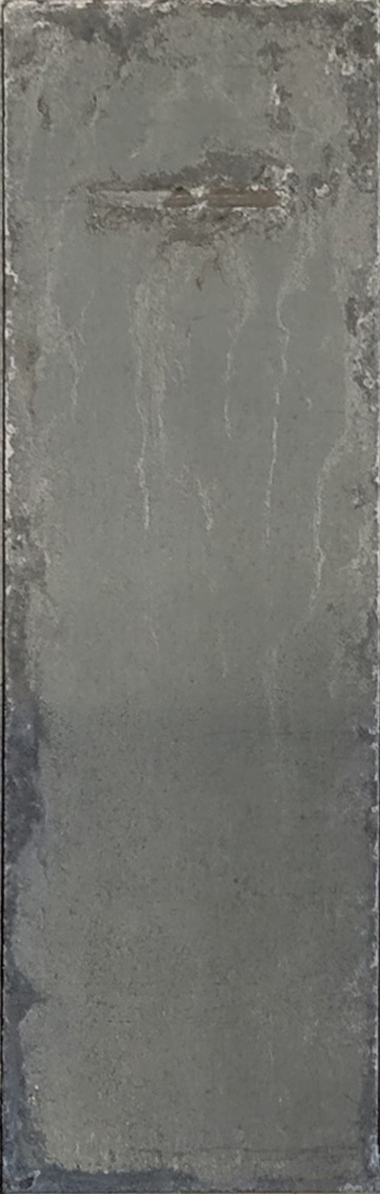} 
        \label{fig:s5_rgb}
                \centerline{(b) Sample five}

    \end{minipage}\hfill
    \begin{minipage}{0.32\linewidth}
        \centering
        \includegraphics[width=\linewidth,height = 8cm]{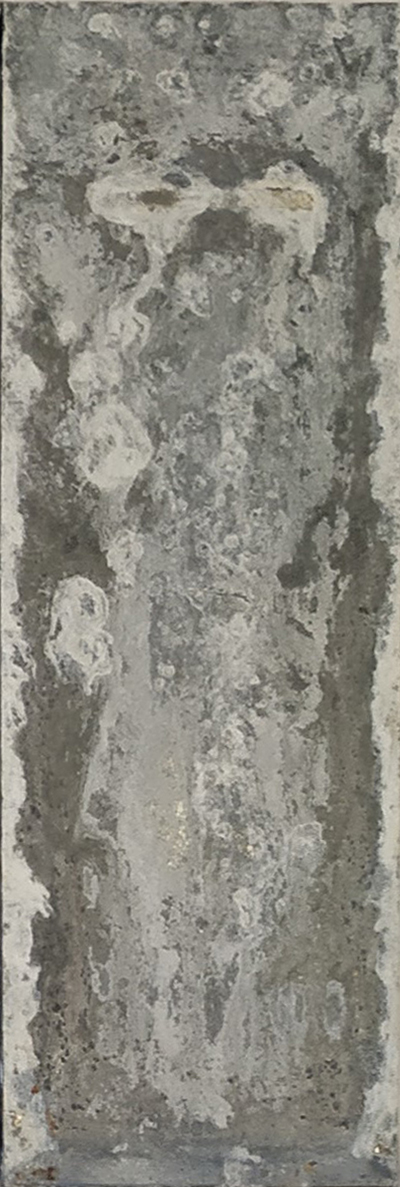} 
        \centerline{(c) Sample ten}
    \end{minipage}

\caption{RGB images of three different samples shown with different salt spray exposure times: (a) Sample one: 24 hours, (b) Sample five: 168 hours and (c) Sample ten: 672 hours}
\label{fig:rgbsamples}
\end{figure}

Ten carbon steel samples with a protective electroplated zinc coating (DX51D-Z275, according to the standard BS EN 10346:2015) were exposed to a salt spray test chamber to accelerate corrosion growth. The samples have a dimension of 150 mm x 50 mm x 1 mm. A 2 mm x 20 mm slot was milled into these samples to expose the bare metal. The salt spray test was performed according to the ISO 9227 standard and each sample was subjected to a different exposure time: 24, 48, 72, 96, 168, 240, 336, 408, 504 and 672 hours. Each sample was rinsed with demineralised water and air dried before the hyperspectral and FTIR measurements. Sample one, five and ten can be seen in Figure \ref{fig:rgbsamples}.

\subsection{Microscope FTIR measurements}
FTIR measurements were performed using a BRUKER Lumos FTIR microscope. Reflectance spectra were recorded in the range from 600 $cm^{-1}$ to 4000 $cm^{-1}$. For one measurement 64 spectra were recorded and averaged. The resolution of the obtained spectra is 4 $cm^{-1}$. For each location, a cluster of four to six points located in close proximity to each other was measured individually. The motorized translation stage is able to record the location of the measurements and an RGB image was taken with the built-in microscope (8x optical zoom). The instrument is calibrated with a stainless steel reference sample. A total of 188 individual points in 36 clusters were measured. The spectra were smoothed and baseline correction was performed using the Spectragryph software \cite{Spectragryph}.

\subsection{Hyperspectral measurements}
Hyperspectral measurements were performed using a push-broom shortwave infrared hyperspectral imaging system. This setup consists of a SWIR camera (SPECIM FX17) mounted above a translation stage (SPECIM LabScanner) with adjustable scanning speed. During acquisition, the samples move while the camera is stationary. The camera can acquire a maximum of 224 bands in the range of 900 nm to 1700 nm for 640 pixels simultaneously. A spatial resolution of 0.17 mm/pixel was obtained. For each measurement, a white reference (with a Spectralon tile) and a dark reference (when the shutter was closed) was recorded. Using these reference measurements, we can calculate the calibrated hyperspectral reflectance image.

\subsection{Hyperspectral classification}
To correlate the hyperspectral and FTIR measurements, we use a two-step process. First, from the FTIR measurements, we identify the most abundant mineral based on spectral features as described in the literature \cite{Kasperek1998,Lebrini2009,Zhu2001,Winiarski2018}. These corrosion minerals are identified as \ch{ZnO} (zincite/zinc oxide) \ch{Zn_{5}(OH)_{8}Cl_{2}.H_{2}O} (simonkolleite), \ch{ZnCO_{3}} (smithsonite), \ch{Zn_{5}(CO _{3})_{2}(OH)_{6}} (marionite/hydrozincite). Resulting in a location in the sample where the identified mineral is present.
Second, using these locations, we can obtain the ground truth spectra in the SWIR range. The SWIR spectra are then used to classify the entire image.
A Savitsky-Golay filter with a window size of 25 points and a second order polynomial is applied to the spectra to smooth out any irregularities for each spectrum. To minimize noise, the first and last 10 bands were excluded from the analysis as the sensor sensitivity is lower in that range. To create a classification map of the entire sample surface, the spectral angle mapping (SAM)\cite{kruse1993spectral} algorithm is used. This algorithm is a measure of how closely the measured spectra are correlated with the reference spectra. Each pixel is compared to the various reference spectra in the five categories (four minerals and the spectra of uncorroded galvanized steel), resulting in a spectral angle for each of the categories, for a single measured spectrum. When applied to the entire image, this results in a classification map in which each pixel is labeled. If the value of the spectral angle is below a certain threshold for a single spectrum, the pixel is classified as Unknown.

\section{Results}
\label{sec:pagestyle}

\subsection{Identifying corrosion minerals using microscopy FTIR}
\begin{figure}[h!tpb]
    \begin{minipage}{\linewidth}
        \centering
        \includegraphics[width=\linewidth]{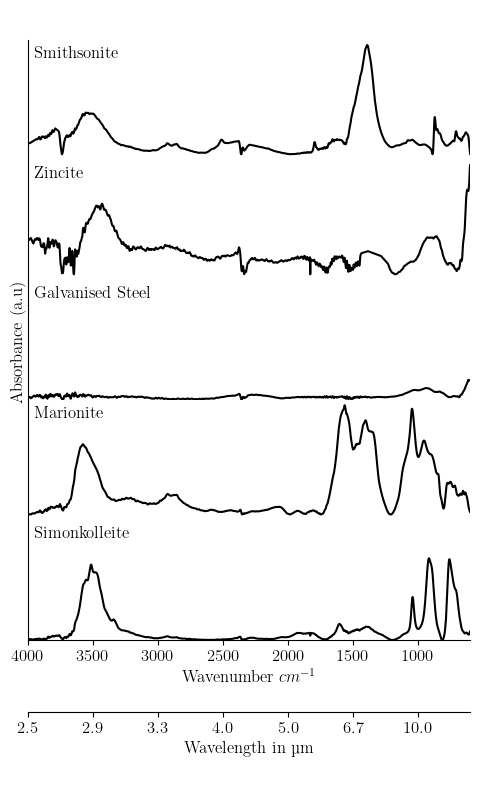} 
        \label{fig:FTIR_spectra}
        \centerline{(a)}

    \end{minipage}\hfill
        \begin{minipage}{\linewidth}
        \centering
        \includegraphics[width=\linewidth]{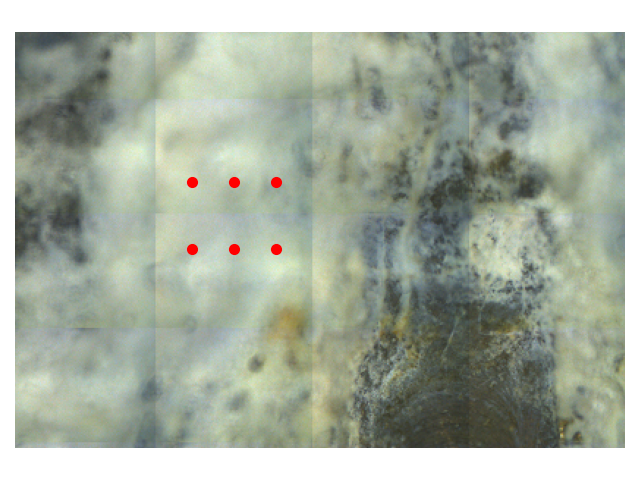} 
        \label{fig:ftirrgb}
                \centerline{(b)}
    \end{minipage}\hfill
    \caption{(a) FTIR spectra of the different corrosion minerals (b) Zoomed RGB image of the microscope FTIR of sample 10. The red dots indicate a single cluster with six separate measurements. The average spectrum of these six points is identified as marionite.}
    \label{fig:FTIRcombo}
\end{figure}

The specific spectra found for each corrosion mineral can be found in Figure \ref{fig:FTIR_spectra}. These spectra are identified by comparing the distinct peaks in the reference spectra with the measured FTIR spectra. Several spectra showed a mixture of two or more minerals, this is evident by the appearance of features for different minerals in a single measured spectrum. These mixed spectra were discarded and only the spectra that show the pure minerals, are used. 

\subsection{Hyperspectral classification}
\begin{figure}[h!]
        \centering
        \includegraphics[width=\linewidth]{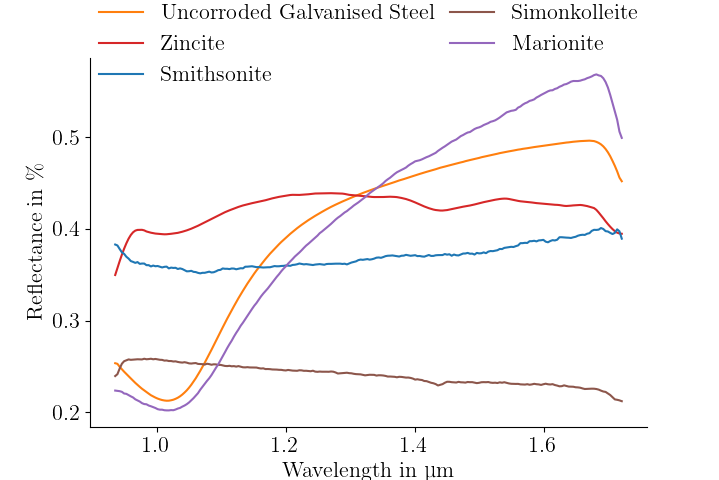}
        \caption{The spectra of the different corrosion minerals.}
\label{fig:hsiSpec}
\end{figure}

From the previous section, we determined the location of the pure minerals using the FTIR measurements on the microscope. The SWIR spectra of the various corrosion minerals can be seen in Figure \ref{fig:hsiSpec}. Several observations can be made from this graph. First, the spectra of marionite and uncorroded electrogalvanized steel are very similar. This can lead to difficulties in distinguishing between these two categories. Second, the spectrum of zincite shows a very distinct low point in the 1440 nm region, while the other spectra do not show this feature. Third, both smithsonite and simonkolleite show very few distinct spectral features. However, the slope and total reflectance values are different.

\begin{figure}[h!tpb]
    \begin{minipage}{\linewidth}
        \centering
        \includegraphics[width=\linewidth]{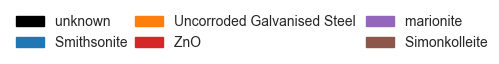} 
    \end{minipage}
    \begin{minipage}{0.32\linewidth}
        \centering
        \includegraphics[width=\linewidth,height = 8cm]{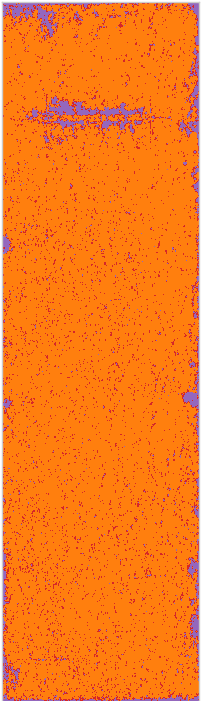} 
        \label{fig:amaps1}
        \centerline{(a)}

    \end{minipage}\hfill
        \begin{minipage}{0.32\linewidth}
        \centering
        \includegraphics[width=\linewidth,height = 8cm]{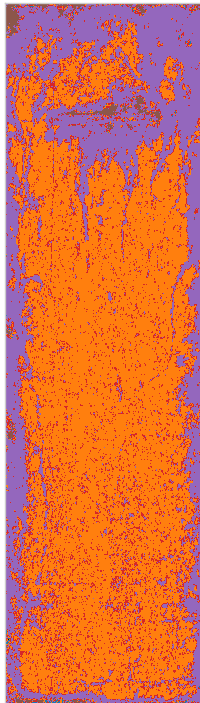} 
        \label{fig:amaps5}
                \centerline{(b)}

    \end{minipage}\hfill
    \begin{minipage}{0.32\linewidth}
        \centering
        \includegraphics[width=\linewidth,height = 8cm]{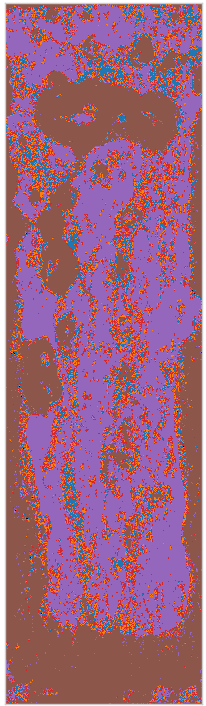} 
        \label{fig:amaps10}
        \centerline{(c)}
    \end{minipage}%
\caption{Three different samples shown with different salt spray exposure times: (a) Sample one: 24 hours, (b) Sample five: 168 hours and (c) Sample ten: 672 hours}
\label{fig:amapsples}
\end{figure}

Looking at the classification maps, seen in Figure \ref{fig:amapsples}, for samples one, five, and ten, we see a gradual increase in the type of minerals present. In sample one, there are only 2 major components: uncorroded galvanized steel and marionite. The mineral marionite is the first corrosion product that has formed, especially around the center cut and around the edges. Sample five still contains a large amount of uncorroded electrogalvanized steel, but the amount of marionite is greater and, in addition, simonkolleite has formed around the edges. Sample ten appears to be completely corroded as there is little evidence of non-corroded galvanized steel. Simonkolleite has formed around the edges and the central cut area. The central portion contains mainly marionite with small patches of zincite occurring at the interface between marionite and simonkolleite. Small amounts of zinc oxide also form, especially around the areas containing zincite.
The other seven samples were also processed and a summary of the classified mineral compositions can be found in the table: \ref{tab:results}
Note that the position of the sample in the salt spray chamber or small impurities in the zinc coating could have an effect on the formation of the various corrosion minerals. Therefore, the mineral abundances do not always increase in a linear way. However, it is possible to see the overall trend for each mineral.

\begin{table*}[h!tpb]
\centering
\begin{tabular}{cccccc}
\multicolumn{1}{l}{} & \multicolumn{5}{c}{\textbf{Mineral classification (\%)}} \\
\textbf{\begin{tabular}[c]{@{}c@{}}Sample \\ number\end{tabular}} & \begin{tabular}[c]{@{}c@{}}Uncorroded \\ Galvananised \\ Steel\end{tabular} & Simonkolleite & Smithsonite & Zincite & Marionite \\
1 & 91.08 & 0.02 & 0.01 & 0.05 & 8.84 \\
2 & 80.76 & 0.04 & 0.19 & 0.02 & 18.97 \\
3 & 76.55 & 0.29 & 0.29 & 0.05 & 22.8 \\
4 & 81.23 & 0.20 & 0.20 & 0.05 & 18.31 \\
5 & 64.24 & 0.77 & 0.69 & 0.18 & 34.09 \\
6 & 48.79 & 7.76 & 0.58 & 0.29 & 41.27 \\
7 & 29.51 & 12,20 & 1.35 & 1.10 & 53.66 \\
8 & 9.83 & 20.32 & 4.58 & 1.49 & 63.58 \\
9 & 2.05 & 35.80 & 6.05 & 1.88 & 54.13 \\
10 & 1.85 & 35.69 & 14.41 & 2.41 & 45.26
\end{tabular}
\caption{The classification results, shown for each category and per sample.}
\label{tab:results}
\end{table*}

\section{Conclusion}
This article presents an alternative chemical imaging method for the detection of corrosion products on zinc electroplated steel by using hyperspectral imaging. An FTIR microscope was used to identify the locations of the pure corrosion minerals (through their spectra). Based on these locations, the specific minerals can be found in the SWIR region. Using the SAM algorithm, we calculate classification maps for each sample with the identified corrosion minerals. This method for identifying corrosion on zinc electroplated steel is faster to other chemical techniques such as FTIR, XRD or SEM. Due to the line scan method, we obtain 640 spectra at once, a remarkable increase compared to the 20 seconds it takes for a single FTIR measurement. This technology can also be applicable outside laboratory conditions. Compared to RGB cameras, this method removes the visual ambiguity between corrosion minerals that appear quite similar in the visible spectrum but are more distinct in the SWIR spectrum. 
Future work could include the use of other techniques, such as XRD measurements or SEM, to validate the results obtained in this work. 

\section{Acknowledgements}
We thank Yanou Fishel for assistance with microscopy FTIR measurements, and Dries Van Hoegaerden for assistance in creating samples and HSI measurements.

\vfill
\pagebreak

\bibliographystyle{IEEEbib}
\bibliography{refs}

\end{document}